 \documentclass[
twocolumn,showpacs, amsmath, amssymb, floatfix,
nofootinbib,wrapfloat byrevtex]{revtex4}

 \usepackage{amsmath,mathrsfs,graphicx,epstopdf,placeins,float}
 \usepackage{booktabs}
 \usepackage{multirow}

 \newcommand{\beq}[1]{\begin{equation}\label{#1}}
 \newcommand{\eeq}{\end{equation}}
 \newcommand{\bea}[1]{\begin{eqnarray}\label{#1}}
 \newcommand{\eea}{\end{eqnarray}}
 \newcommand\figcaption{\def\@captype{figure}\caption}
 \newcommand\tabcaption{\def\@captype{table}\caption}

\begin{document}

 \title{An inconsistency in the spectrum of bosonic open 2-brane}
 \author{M. Abdul Wasay $^{\!\!1,2}$}
\email{wasay31@gmail.com}

\author{Ding-fang Zeng$^{1,}$}
\email{dfzeng@bjut.edu.cn}
\affiliation{$^1$Institute of Theoretical Physics, Beijing University of Technology, Beijing 100124, China\\
$^2$Department of Physics, University of Agriculture, Faisalabad, Pakistan}
 \begin{abstract}
 \section*{Abstract}
 We show that the spectrum of a bosonic open 2-brane does not contain any massless states to take the role of gravitons. Moreover, the spectrum of this open 2-brane only contains half integer mass squared values.
  \end{abstract}
 \pacs{11.25.-w,~12.60.Jv}

 \maketitle
 \smallskip

\section{Introduction}

Besides the impressive progress of string theories in the past few decades which compelled physicists to believe that a consistent theory of quantum gravity based on string philosophy would be built soon has slowed down. The early success of string theory also sparked the motivation to study higher dimensional extended objects. Indeed if one can give up 0 dimensional particles in favor of strings which are one dimensional objects, then why not strings in favor of two dimensional membranes?
The basic idea that elementary particles could be interpreted as vibrating modes of a membrane originally came in 1962 by Dirac~\cite{Dirac}.

Since the task of achieving a consistent theory of quantum gravity based on string philosophy became more and more subtle, physicists tried to find alternative candidates to construct a theory of fundamental interactions, one of these candidates was supermembrane theory~\cite{supermembrane1}\cite{supermembrane2}\cite{supermembrane3}. However, it was soon realized that the supermembrane has an unstable ground state~\cite{supermembrane}.

Compared to the works on string quantization, works on membrane quantization are very few, precisely because of the difficulty the Nambu action meets when one tries to generalize it to $2+1$ dimensions or general $p+1$ dimensions. However, it is possible to quantize a bosonic membrane theory in a flat background~\cite{open2brane}\cite{open2brane1}.

We will analyze the mass squared values in the spectrum of the bosonic open 2-branes. The bosonic membranes could be related to the bosonic
string theory via dimensional reduction~\cite{Membrane}. The critical spacetime dimension in analogy with bosonic string theory is taken to be $D=26$. We will focus on the spectrum of states after reviewing their quantization.

The organization of this note is: Section II is a brief review of the quantization scheme in a flat background~\cite{open2brane}. Section III deals with the spectrum of this open 2-brane and section IV is summary and conclusion.
\section{A Brief Review}
The quantization of bosonic open branes was studied in references \cite{openpbrane}\cite{open2brane}\cite{open2brane1}.
The open 2-brane dynamics is captured by the Nambu-Goto action, which physically describes the world-volume swept out by the brane. However, for quantization purposes it is not suitable because of the square root. However, one may follow reference \cite{Banarjee} and write a Polyakov-type action for the general bosonic open 2-brane given by
\bea{}
S=-\frac{1}{4\pi\alpha^\prime}\int d^3\sigma\sqrt{-h}[h^{\alpha\beta}\partial_\alpha X^\mu\partial_\beta X_\mu+R-2\Lambda ]
\label{theaction}
\eea
where $R$ and $\Lambda$ give the contribution from the cosmological constant term. A 2-brane sweeps out a 3-dimensional world-volume which is parameterized by $\tau, \sigma^1, \sigma^2$.

The energy-momentum tensor is given by the variation of the action \eqref{theaction}, with respect to the world-volume metric $h^{\alpha\beta}$, as
\bea{}
T_{\alpha\beta}&=&\frac{-2\pi\alpha^\prime}{\sqrt{-h}}\frac{\delta S}{\delta
h^{\alpha\beta}}
\\ \nonumber
&=&\partial_\alpha X^\mu\partial_\beta X_\mu-\frac{1}{2}h_{\alpha\beta}h^{\gamma\delta}\partial_\gamma X^\mu\partial_\delta X_\mu+\frac{1}{2}h_{\alpha\beta}
\\
&=&0 \nonumber
\label{tensor}
\eea

Under the flat metric condition $h_{\alpha\beta}=\eta_{\alpha\beta}$, the components of this energy-momentum tensor can be written as
\bea{}
T_{\alpha\beta}=\partial_\alpha X^\mu\partial_\beta X_\mu-\frac{1}{2}h_{\alpha\beta}[-\partial_\tau X^\mu \partial_\tau X_\mu+~~~~\nonumber
\\
\partial_{\sigma^1} X^\mu \partial_{\sigma^1} X_\mu +
\partial_{\sigma^2} X^\mu \partial_{\sigma^2} X_\mu] +\frac{1}{2}h_{\alpha\beta}~\nonumber
\eea
with
\bea{}
T_{00}&=&\frac{1}{2}\left[\partial_\tau X^\mu \partial_\tau X_\mu\!\!+\partial_{\sigma^1} X^\mu \partial_{\sigma^1} X_\mu\!\!+\partial_{\sigma^2} X^\mu \partial_{\sigma^2} X_\mu\!\!-\!1\right]\nonumber
\\
&=&0
\eea
\bea{}
T_{11}&=&\frac{1}{2}\left[\partial_\tau X^\mu \partial_\tau X_\mu\!\!+\partial_{\sigma^1} X^\mu \partial_{\sigma^1} X_\mu\!\!-\partial_{\sigma^2} X^\mu \partial_{\sigma^2} X_\mu\!\!+\!1\right]\nonumber
\\
&=&0
\eea
and
\bea{}
T_{22}&=&\frac{1}{2}\left[\partial_\tau X^\mu \partial_\tau X_\mu\!\!-\partial_{\sigma^1} X^\mu \partial_{\sigma^1} X_\mu\!\!+\partial_{\sigma^2} X^\mu \partial_{\sigma^2} X_\mu\!\!+\!1\right]\nonumber
\\
&=&0
\eea
and the Euler-Lagrange equation for the $X^\mu$ fields is given by
\bea{}
(\partial_\tau^2-\partial_{\sigma_1}^2-\partial_{\sigma_2}^2)X^\mu(\tau,\sigma_1,\sigma_2)=0
\label{ELequation2brane}
\eea
By imposing Neumann boundary condition
\bea{}
\partial_{\sigma^1}X^\mu(\tau,0,\sigma^2)=\partial_{\sigma^1}X^\mu(\tau,\pi,\sigma^2)=0
\\
\partial_{\sigma^2}X^\mu(\tau,\sigma^1,0)=\partial_{\sigma^2}X^\mu(\tau,\sigma^1,\pi)=0
\eea
one can get the following modes expansion for the $X^\mu$ fields

\bea{}
X^\mu(\tau,\sigma^1,\sigma^2)=\frac{x^\mu}{\sqrt\pi}\!+\!\frac{2\acute{\alpha}p^\mu}{\sqrt\pi}\tau\!+\!i\sqrt{2\acute{\alpha}}\sum\limits _{m,n=0}^{+\infty}\!\!\!(n^2+m^2)^{-\frac{1}{4}}
\\
\times
\left(X^\mu_{nm}e^{i\tau\sqrt{n^2+m^2}}-X^{\dag \mu}_{nm}e^{-i\tau\sqrt{n^2+m^2}} \right)
\nonumber
\\
\times\textmd{cosn}\sigma^1\textmd{cosm}\sigma^2\nonumber
\eea
and the canonical momentum, as
\bea{}
P^\mu(\sigma)=\frac{p^\mu}{\pi\sqrt\pi}+\frac{1}{\pi}\sqrt{\frac{2}{\acute{\alpha}}}\sum\limits _{m,n=0}^{+\infty}(n^2+m^2)^{\frac{1}{4}}
\\ \times
\left(P^{\mu\dag}_{nm}e^{i\tau\sqrt{n^2+m^2}}+P^\mu_{nm}e^{-i\tau\sqrt{n^2+m^2}} \right)\nonumber\\ \times\textmd{cosn}\sigma^1\textmd{cosm}\sigma^2
\nonumber
\eea

According to the standard commutation relation
\bea{}
\left[X^\mu,P^\nu\right]
=\eta^{\mu\nu}\delta(\sigma^1-\sigma^{\prime1})\delta(\sigma^2-\sigma^{\prime2})
\label{commutationrelation1}
\eea
the following commutation relations between the creation/annihilation operators can be obtained
\bea{}
\left[X^\mu_{nm},P^\nu_{n^\prime m^\prime} \right]=\frac{1}{\pi}\eta^{\mu\nu}[\delta_{nn^\prime}\delta_{mm^\prime}-\frac{1}{2}\delta_{-n,n^\prime}\delta_{mm^\prime}\nonumber
\\
-\frac{1}{2}\delta_{nn^\prime}\delta_{-m,m^\prime} ]
\eea
Using these creation/annihilation operators, the Hamiltonian of the system is expressed as
\bea{}
4\pi\alpha^\prime H=\int\limits_0^\pi d\sigma^1\int\limits_0^\pi d\sigma^2\left(P_\mu \dot X^\mu-\mathcal L \right)~~~~~~~~~~~
\\
=2\alpha^\prime\pi^2\eta_{\mu\nu}\sum\limits_{n=1}^\infty n\left[X^{\mu\dag}_{n0}X^\nu_{n0}+X^\mu_{n0}X^{\nu\dag}_{n0}\right]
\label{hamiltonianof2brane}\nonumber\\
+2\alpha^\prime\pi^2\eta_{\mu\nu}\sum\limits_{m=1}^\infty m\left[X^{\mu\dag}_{0m}X^\nu_{0m}+X^\mu_{0m}X^{\nu\dag}_{0m}\right]
\nonumber\\
+\alpha^\prime\pi^2\eta_{\mu\nu}\sum\limits_{n,m=1}^\infty(n^2+m^2)^{\frac{1}{2}}\left[X^{\mu\dag}_{nm}X^\nu_{nm}+X^\mu_{nm}X^{\nu\dag}_{nm} \right]
\nonumber\\
+4\pi\alpha^{\prime2}p^2~~~
\eea

Under the tensorial assumption,
\bea{}
X^\mu_{nm}&=&\sqrt{\frac{2}{\pi}}~\phi^{\mu\dag}_n\otimes\phi^{\mu\dag}_m
\\
X^{\mu\dag}_{nm}&=&\sqrt{\frac{2}{\pi}}~\phi^{\mu}_n\otimes\phi^{\mu}_m
\eea
the Hamiltonian \eqref{hamiltonianof2brane} is translated into
\bea{}
H=N_1+N_2+N_{12}+a+b-\alpha^\prime M^2
\label{massoperator}
\eea
where
\bea{}
N_1&=&\eta_{\mu\nu}\sum\limits_{n=1}^\infty n\left\{\phi^{\mu\dag}_n\otimes\phi^{\mu\dag}_0 \right\}\left\{\phi^{\mu}_n\otimes\phi^{\mu}_0 \right\}
\\
N_2&=&\eta_{\mu\nu}\sum\limits_{m=1}^\infty m\left\{\phi^{\mu\dag}_0\otimes\phi^{\mu\dag}_m \right\}\left\{\phi^{\mu}_0\otimes\phi^{\mu}_m \right\}
\\
N_{12}&=&\eta_{\mu\nu}\!\!\!\!\sum\limits_{n,m=1}^\infty\!\!\!(n^2\!+\!m^2)^{\frac{1}{2}}\left\{\phi^{\mu\dag}_n\otimes\phi^{\mu\dag}_m \right\}\left\{\phi^{\mu}_n\otimes\phi^{\mu}_m \right\}
\\
a&=&\eta^\mu_\mu\sum\limits_{n=1}^\infty n
 \label{a}
\\
b&=&\frac{1}{2}\eta^\mu_\mu\sum\limits_{n,m=1}^\infty\sqrt{n^2+m^2}
\label{b}
\eea

\section{Spectrum}
We will employ the Zeta function regularization scheme to regularize the infinite divergent summation in \eqref{a} and \eqref{b}. The contributions arising from the infinite sum \eqref{b} have not been considered carefully in earlier works and have usually been taken as some background field effects.

In fact, there is another scheme \cite{zetaRegularization}, the so called Epstein Zeta functions to regularize the infinite sums of type \eqref{b}.
\bea{}
&&\sum\limits_{n,m=1}^\infty\sqrt{\left(\frac{n}{c_1}\right)^2+\left(\frac{m}{c_2}\right)^2}
\nonumber\\
&&=\frac{1}{24}\left(\frac{1}{c_1}+\frac{1}{c_2}\right)-\frac{\zeta(3)}{8\pi^2}\left(\frac{c_1}{c_2^2}+\frac{c_2}{c_1^2}\right)
\\
&&-\frac{\pi^{\frac{3}{2}}}{2\sqrt{c_1c_2}}\left[\textmd{exp}\left(-2\pi\frac{c_1}{c_2}\right)\left(1+\mathcal{O}(10^{-3}) \right) \right]
\nonumber
\eea
with $\zeta(3)\approx1.202$ and $c_1\leq c_2$. The infinite sums \eqref{a} and \eqref{b} appearing in the quantization of the open-2 brane, can be written as
\bea{}
a&=&\eta^\mu_\mu\sum\limits_{n=1}^\infty n=(D-2)\sum\limits_{n=1}^\infty n = (D-2)\zeta(-1)
\eea
\bea{}
b=\frac{1}{2}\eta^\mu_\mu\sum\limits_{n,m=1}^\infty\!\!\!\sqrt{n^2+m^2}=\frac{D-2}{2}\sum\limits_{n,m=1}^\infty\!\!\!\sqrt{n^2+m^2}
\eea
Numerically, $a=-2$ while $b\approx\frac{1}{2}$, given that the spacetime dimension $D=26$.

According to the above calculation and the fact that $T_{00}=0=H$, we can rewrite Eq. \eqref{massoperator}, the mass formula of the open 2-brane spectrum , as
\bea{}
\alpha^\prime M^2=N_1+N_2+N_{12}-\frac{3}{2}
\eea
By definition, the ground state of the system is annihilated by the operators $\phi^\mu_{n\geqslant0}$ and $\phi^\mu_{m\geqslant0}$, with the condition that $n\cdot m\neq0$
\bea{}
\phi_n\otimes\phi_m|0,0\rangle=0
\eea
A general state $|\varphi,\chi\rangle$, in the Fock space can be obtained by application of the creation operators $\phi^{\mu\dag}_n$ and $\phi^{\mu\dag}_m$ on the ground state
\bea{}
|\varphi,\chi\rangle=\Pi_{n,m}\left\{\phi^{\mu\dag}_{n_i}\otimes\phi^{\mu\dag}_{m_i} \right\}|0,0\rangle
\eea
with $\sum_i n_i=\varphi$ and $\sum_i m_i=\chi$. The states $|\varphi,\chi\rangle$ and $|\chi,\varphi\rangle$, have the same mass squared.

At the lowest mass level, the number operators $N_1$, $N_2$ and $N_{12}$ are zero, so
\bea{}
\alpha^\prime M^2=-\frac{3}{2}.
\eea
The first excited level contains two kinds of states corresponding to $n=1$, $m=0$ and $n=0$, $m=1$, i.e.
\bea{}
|1,0\rangle&=&k_\mu\left(\phi^{\mu\dag}_1\otimes\phi^{\mu\dag}_0 \right)|0,0\rangle
\\ \nonumber
\\
|0,1\rangle&=&l_\mu\left(\phi^{\mu\dag}_0\otimes\phi^{\mu\dag}_1 \right)|0,0\rangle
\eea
whose mass-square operator reads
\bea{}
\alpha^\prime M_k^2&=&N_1+N_2+N_{12}-\frac{3}{2} \nonumber
\\
&=&1+0+0-\frac{3}{2} \nonumber
\\
&=&-\frac{1}{2}
\eea
\bea{}
\alpha^\prime M_l^2&=&N_1+N_2+N_{12}-\frac{3}{2} \nonumber
\\
&=&0+1+0-\frac{3}{2} \nonumber
\\
&=&-\frac{1}{2}
\eea
The ground and first excited states are both tachyonic. However the value of mass-squared operator in~\cite{open2brane} is an integer, while we show that it is half integer valued.
The ground states are scalar states.
We have two different kind of vector states at the first excited state level corresponding to $|1,0\rangle$ and $|0,1\rangle$.

At the second excited level, there are four kinds of tensor states, featured by
\bea{}
\begin{cases}
n=0\\
m=2
\end{cases} \qquad, \qquad \begin{cases}
n=2\\
m=0
\end{cases}
\eea
and
\bea{}
\begin{cases}
n_1=n_2=1\\
m_1=m_2=0
\end{cases} \qquad, \qquad \begin{cases}
n_1=n_2=0\\
m_1=m_2=1
\end{cases}
\eea
with the creation operators arranging at this level as follows
\bea{}
|2,0\rangle_1&=&k_{\mu\nu}\left(\phi^{\mu\dag}_1\otimes\phi^{\mu\dag}_0 \right)\left(\phi^{\nu\dag}_1\otimes\phi^{\nu\dag}_0 \right)|0,0\rangle\nonumber
\\ \nonumber
\\
|2,0\rangle_2&=&k_\mu\left(\phi^{\nu\dag}_2\otimes\phi^{\nu\dag}_0 \right)|0,0\rangle\nonumber
\\ \nonumber
\\
|0,2\rangle_1&=&l_{\mu\nu}\left(\phi^{\mu\dag}_0\otimes\phi^{\mu\dag}_1 \right)\left(\phi^{\nu\dag}_0\otimes\phi^{\nu\dag}_1 \right)|0,0\rangle\nonumber
\\ \nonumber
\\
|0,2\rangle_2&=&l_\mu\left(\phi^{\nu\dag}_0\otimes\phi^{\nu\dag}_2 \right)|0,0\rangle\nonumber
\eea
For these states, the number operators are $N_1=2$, $N_2=N_{12}=0$ or $N_2=2$, $N_1=N_{12}=0$. The mass squared operator will become
\bea{}
\alpha^\prime M^2=&=&N_1+N_2+N_{12}-\frac{3}{2} \nonumber
\\ \nonumber
\\
&=&0+2+0-\frac{3}{2}~~ \textmd{or}~~ 2+0+0-\frac{3}{2} \nonumber
\\ \nonumber
\\
&=&\frac{1}{2}
\eea
We can see that the spectrum of the bosonic 2-brane does not contain any massless states to play the role of gravitons. Moreover, the mass squared operator for the open 2-brane only gives half integer mass squared values.

\section{Summary and Conclusion}

In this paper, firstly we review the quantization of open 2-branes, starting from the Polyakov action. We then investigate the spectrum of the open 2-brane by taking into account the contributions of both the infinite sums \eqref{a} and \eqref{b}. On the basis of this we can draw the conclusion that there are no gravitons present in the bosonic open 2-brane at the massless level, or there are no massless states in the bosonic open 2-brane spectrum. This further implies that the bosonic open 2-brane theory is not a theory of quantum gravity. This is reminiscent of the fact that in the string case \cite{Polchinski}, simple open strings do not form self-complete system of quantum gravity.

\section*{Acknowledgments}

We thank Jian-Feng Wu for suggesting us literatures on the special zeta functions \cite{zetaRegularization}. The work is supported by the Open Project Program of State Key Laboratory of Theoretical Physics, Institute of Theoretical Physics, Chinese Academy of Sciences, China. M.A.Wasay would also like to thank Yasir Jamil for providing the facilities at Physics department University of Agriculture, Faisalabad, where part of this work was carried out.

\section*{Conflict of Interests}

The authors declare that there is no conflict of interests regarding the publication of this paper.

\end{document}